\documentclass{aa}  

\usepackage{graphicx}
\usepackage{txfonts}
\DeclareRobustCommand{\VAN}[3]{#2}
\let\VANthebibliography\thebibliography
\def\thebibliography{\DeclareRobustCommand{\VAN}[3]{##3}\VANthebibliography}

\usepackage{color}

\usepackage{textcomp}
\usepackage{amsmath}	% Advanced maths commands
\usepackage{amssymb}	% Extra maths symbols
\begin{document}

   \title{Very Large Array observations of the mini-halo and AGN feedback in the Phoenix cluster}
   \titlerunning{VLA observations of the mini-halo and AGN feedback in the Phoenix cluster}

   \author{R. Timmerman
          \inst{1}\fnmsep\thanks{E-mail: rtimmerman@strw.leidenuniv.nl (RT)}
          \and
          R. J. van Weeren\inst{1}
          \and
          M. McDonald\inst{2}
          \and
          A. Ignesti\inst{3,4}
          \and
          B. R. McNamara\inst{5}
          \and
          J. Hlavacek-Larrondo\inst{6}
          \and
          H. J. A. Röttgering\inst{1}
          }

   \institute{Leiden Observatory, Leiden University, P.O. Box 9513, 2300 RA Leiden, The Netherlands
            \and
            Kavli Institute for Astrophysics and Space Research, Massachusetts Institute of Technology, 77 Massachusetts Avenue, Cambridge, MA 02139, USA
            \and
            DIFA, University of Bologna, Via Gobetti 93/2, 40129 Bologna, Italy
            \and
            IRA INAF, Via Gobetti 101, 40129 Bologna, Italy
            \and
            Department of Physics and Astronomy, University of Waterloo, Waterloo, ON, Canada
            \and
            Département de Physique, Université de Montréal, C.P. 6128, Succ. Centre-Ville, Montréal, Québec H3C 3J7, Canada
            }

   \date{Received XXX; accepted YYY}

% \abstract{}{}{}{}{} 
% 5 {} token are mandatory
 
  \abstract
   {The relaxed cool-core Phoenix cluster (SPT-CL J2344-4243) features an extremely strong cooling flow, as well as a mini-halo. Strong star-formation in the brightest cluster galaxy indicates that AGN feedback has been unable to inhibit this cooling flow.}
   {We have studied the strong cooling flow in the Phoenix cluster by determining the radio properties of the AGN and its lobes. In addition, we use spatially resolved radio observations to investigate the origin of the mini-halo.}
   {We present new multifrequency Very Large Array 1--12~GHz observations of the Phoenix cluster which resolve the AGN and its lobes in all four frequency bands, and resolve the mini-halo in L- and S-band.}
   {Using our L-band observations, we measure the total flux density of the radio lobes at 1.5 GHz to be \(7.6\pm0.8\) mJy, and the flux density of the mini-halo to be \(8.5\pm0.9\) mJy. Using high-resolution images in L- and X-band, we produce the first spectral index maps of the lobes from the AGN and measure the spectral indices of the northern and southern lobes to be -1.35 \(\pm\) 0.07 and -1.30 \(\pm\) 0.12, respectively. Similarly, using L- and S-band data, we map the spectral index of the mini-halo, and obtain an integrated spectral index of \(\alpha=-0.95 \pm 0.10\).}
   {We find that the mini-halo is most likely formed by turbulent re-acceleration powered by sloshing in the cool core due to a recent merger. In addition, we find that the feedback in the Phoenix cluster is consistent with the picture that stronger cooling flows are to be expected for massive clusters like the Phoenix cluster, as these may feature an underweight supermassive black hole due to their merging history. Strong time variability of the AGN on Myr-timescales may help explain the disconnection between the radio and the X-ray properties of the system. Finally, a small amount of jet precession of the AGN likely contributes to the relatively low ICM re-heating efficiency of the mechanical feedback.}


   \maketitle
%
%-------------------------------------------------------------------

\section{Introduction}

The emission of strong X-ray radiation by the intracluster medium (ICM) in galaxy clusters suggests that this medium should often cool down rapidly, within a time scale of \(\sim 10^9\) years or less \citep[e.g.][]{fabian94}. As the ICM cools down, it is expected to flow down the gravitational well of the cluster, and accrete onto the galaxy at the center. This accretion of matter should then trigger star-formation in the central galaxies proportional to the cooling flow of the ICM. However, both this cooling of the ICM and the star-formation in the center of the cluster are observed to be much weaker than expected \citep{fabian82, mcnamara89, page12, mcdonald18}, leading to what is known as the "cooling flow problem". The generally accepted solution to this problem is that feedback from active galactic nuclei (AGN) supplies energy to the ICM in the form of radiation and jetted outflows of plasma, thereby preventing the medium from cooling down \citep[e.g.][]{bruggen02, mcnamara07, fabian12}.

Studying this feedback process is essential to our understanding of the formation and evolution of galaxies, as it plays a critical role in the cooling of the ICM and the star-formation in galaxies across cosmic time \citep[e.g.][]{dimatteo05, croton06, menci06, sijacki07, lagos08, ciotti10, mathews11, vogelsberger14, rasia15}. In particular, clusters of galaxies form a great opportunity to study AGN feedback, due to the relatively dense ICM, which is capable of creating strong cooling flows. The ICM is often also dense enough to keep the jetted outflows from the AGN contained, which allows this mechanical form of feedback to be studied in detail \citep{mcnamara12}.

The interaction between the ICM and the jetted outflows from the AGN can be directly observed in the X-ray regime. As the jetted outflows displace the ICM, they create large cavities which are observed as depressions in X-ray observations. At radio frequencies, bubbles of synchrotron-emitting plasma are observed to be coincident with these cavities, verifying that these cavities are produced by the AGN \citep{gull73, gitti11}.

Despite much research into AGN feedback in galaxy clusters, many open questions have remained. Of particular interest for this work is the connection between the central AGN and the mini-halo surrounding the brightest cluster galaxy (BCG). Mini-halos are faint diffuse synchrotron emission regions commonly found in relaxed cool-core galaxy clusters. They span a region of a few hundred kpc, as they are generally confined to the cool core region of a galaxy cluster. Mini-halos often feature an amorphous shape, and have been found to show steep spectral indices of around \(\alpha=-1\) to \(\alpha=-1.5\) \citep{gitti04a, giacintucci19, vanweeren19}.

The origin of mini-halos remains a topic of debate, and is hampered by the difficulty of obtaining high-quality data on mini-halos, as the central AGN often dominates the view. The emission of synchrotron radiation in the mini-halo means that there must be a population of cosmic-ray electrons present in a magnetic field. However, based on the short lifetime of these electrons of 10--100 Myr, they must be accelerated in-situ \citep{brunetti14}. Two mechanisms are proposed by which the electrons can be accelerated. In the hadronic model, relativistic, secondary electrons are injected by collisions between relativistic and thermal protons \citep[e.g.][]{pfrommer04, fujita07}. This model predicts the presence of diffuse gamma-ray emission, produced by the same proton-proton collisions and a gradual decrease of the radio emission due to the diffusion of cosmic ray (CR) protons in the ICM. In the case of radio emission produced purely by secondary electrons, the spectral index would depend only on the energy distribution of the CR protons and hence it would not vary with the radius.
Alternatively, according to the turbulent re-acceleration model, fossil electrons from the AGN are re-accelerated to high energies by magneto-hydrodynamic turbulence in the cluster \citep[e.g.][]{gitti02, mazzotta08, zuhone13}. This turbulence is thought to generally be caused by strong cooling flows or a merger event in the recent history of the cluster, although recent observations of a Mpc-scale radio halo in a cool core cluster \citep{bonafede14} and ultra steep-spectrum emission extending beyond the cool core \citep{savini18} have challenged that idea. The turbulent re-acceleration model predicts a possible steepening of the spectrum with radius, and a radio brightness profile which is strongly contained by the cold fronts produced in the ICM by a recent merger event. These cold fronts are density discontinuities created by the cold and dense gas from the cool core or a subcluster moving through the surrounding hot gas, and therefore also commonly form the boundary of a turbulent region \citep[e.g.][]{markevitch07}. Accurately determining the properties of mini-halos is essential to understand the underlying acceleration mechanism, and thereby their origin.

In this paper, we adopt a \(\Lambda\)CMD cosmology, with cosmological parameters of \(H_0\ =\ 70 \mathrm{km}\ \mathrm{s}^{-1}\ \mathrm{Mpc}^{-1}\), \(\Omega_m\ =\ 0.3\) and \(\Omega_\Lambda\ =\ 0.7\). In this cosmology, the luminosity distance to the Phoenix cluster at \(z\ =\ 0.597\) is 3\,508 Mpc and an angular scale of 1 arcsecond at this redshift corresponds to 6.67 kpc. Furthermore, we follow the convention of defining our spectral indices according to \(S \propto \nu^\alpha\).

\section{The Phoenix Cluster}

In this work, we focus on the Phoenix cluster (SPT-CL\,J2344$-$4243), a massive galaxy cluster at redshift \(z=0.597\) discovered by \citet{williamson11} in the 2\,500 deg\(^2\) South Pole Telescope Survey. The Phoenix cluster is a relaxed cool-core cluster featuring a type 2 QSO \citep{ueda13}, and is characterized by its high cooling flow and star-formation rate. X-ray, optical and infrared observations by \citet{mcdonald12, mcdonald13, mcdonald19} show a cooling flow of \(\sim3\,100\ M_\odot\) per year, with a star-formation rate of \(\sim800\ M_\odot\ yr^{-1}\). Whereas the star-formation rate is generally on the order of 1\% of the predicted cooling flow, for the Phoenix cluster this ratio is almost 30\%, indicating that the feedback process has not been able to completely inhibit the cooling flow. Spectroscopic observations of the warm and cold gas in the core of the Phoenix cluster suggest that this rapid star-formation may be a short phase, as the molecular gas supply is expected to deplete on a timescale of \(\sim30\ \mathrm{Myr}\) \citep{mcdonald14}.

Using \textit{Chandra} observations, \citet{larrondo14} revealed the presence of X-ray cavities in the ICM. Follow-up research using deeper observations by \citet{mcdonald15, mcdonald19} provides an estimate of the scale of these cavities of 8--14 kpc, which suggests a jet power from the AGN of \(1.0^{+1.5}_{-0.4}\times10^{46}\) erg s\(^{-1}\). In addition, star-forming filaments extending up to 50--100 kpc from the core of the cluster were observed using deep optical imaging. ALMA observations also show molecular gas filaments with a length of 10--20 kpc tracing the edges of the X-ray cavities \citep{russell17}.

Observations using the \textit{Giant Metrewave Radio Telescope} (GMRT) by \citet{vanweeren14} uncovered a mini-halo surrounding the BCG, which extends a region of 400--500 kpc. This mini-halo was later observed by \citet{raja20} using the \textit{Karl G. Jansky Very Large Array} (VLA) in CnB-configuration. By subtracting compact emission from their data, they detect the mini-halo and derive a flux density of the mini-halo at 1.5 GHz of 9.65 \(\pm\) 0.97 mJy. They find that the 3\(\sigma\) contours of their map span a region of 310 kpc.

New deep \textit{Chandra} and X-band VLA observations by \citet{mcdonald19} have revealed radio jets that are coincident with the previously detected X-ray cavities. In addition, they present \textit{Hubble} observations suggesting that the AGN may lift cool low-entropy gas up to larger radii where it can cool faster than its fallback time, resulting in multiphase condensation. The gas kinematics and strong high-ionization emission lines indicate that relatively strong turbulence may be present in the core.

In this paper, we aim to investigate the strong cooling flow observed in the Phoenix cluster by imaging the AGN and its jetted outflows across a wide range of radio frequencies. In addition, we aim to study the origin of the mini-halo by measuring its properties using spatially resolved radio observations for the first time.

\section{Observations and data reduction}

\begin{table}
    \centering\small
    \caption{Summary of the observations}
    \begin{tabular}{lllll}
        \hline\hline
        Configuration & Obs. date & Freq. & Int. time & \(\theta_\text{FWHM}\)\\
         & & (GHz) & (s) & (")\\\hline
        L-band, A-array & 24 Mar. 2018 & 1-2 & 2 & 3 \(\times\) 0.7\\
        L-band, B-array & 2 Nov. 2017 & 1-2 & 3 & 10 \(\times\) 2\\
        L-band, CnB-array& 23 Jan. 2015 & 1-2 & 5 & 16 \(\times\) 9\\
        \small\textit{(archival)} \\ \hline
        S-band, A-array & 11 Mar. 2018 & 2-4 & 2 & 2 \(\times\) 0.4\\
        S-band, B-array & 25 Sep. 2017 & 2-4 & 3 & 6 \(\times\) 1\\
        S-band, C-array & 10 Jun. 2017 & 2-4 & 5 & 18 \(\times\) 4\\\hline
        C-band, A-array & 4 Mar. 2018 & 4-8 & 2 & 1 \(\times\) 0.3\\
        C-band, B-array & 8 Sep. 2017 & 4-8 & 3 & 3 \(\times\) 0.6\\
        C-band, C-array & 9 Jun. 2017 & 4-8 & 5 & 9 \(\times\) 2\\\hline
        X-band, A-array & 6 Mar. 2018 & 8-12 & 2 & 0.5 \(\times\) 0.1\\
        X-band, B-array & 7 Sep. 2017 & 8-12 & 3 & 2 \(\times\) 0.4\\
        X-band, C-array & 5 Jun. 2017 & 8-12 & 3 & 6 \(\times\) 1\\\hline\hline
    \end{tabular}
    \label{tab:obs}
\end{table}

\begin{figure*}
    \centering
    \includegraphics[width=\textwidth]{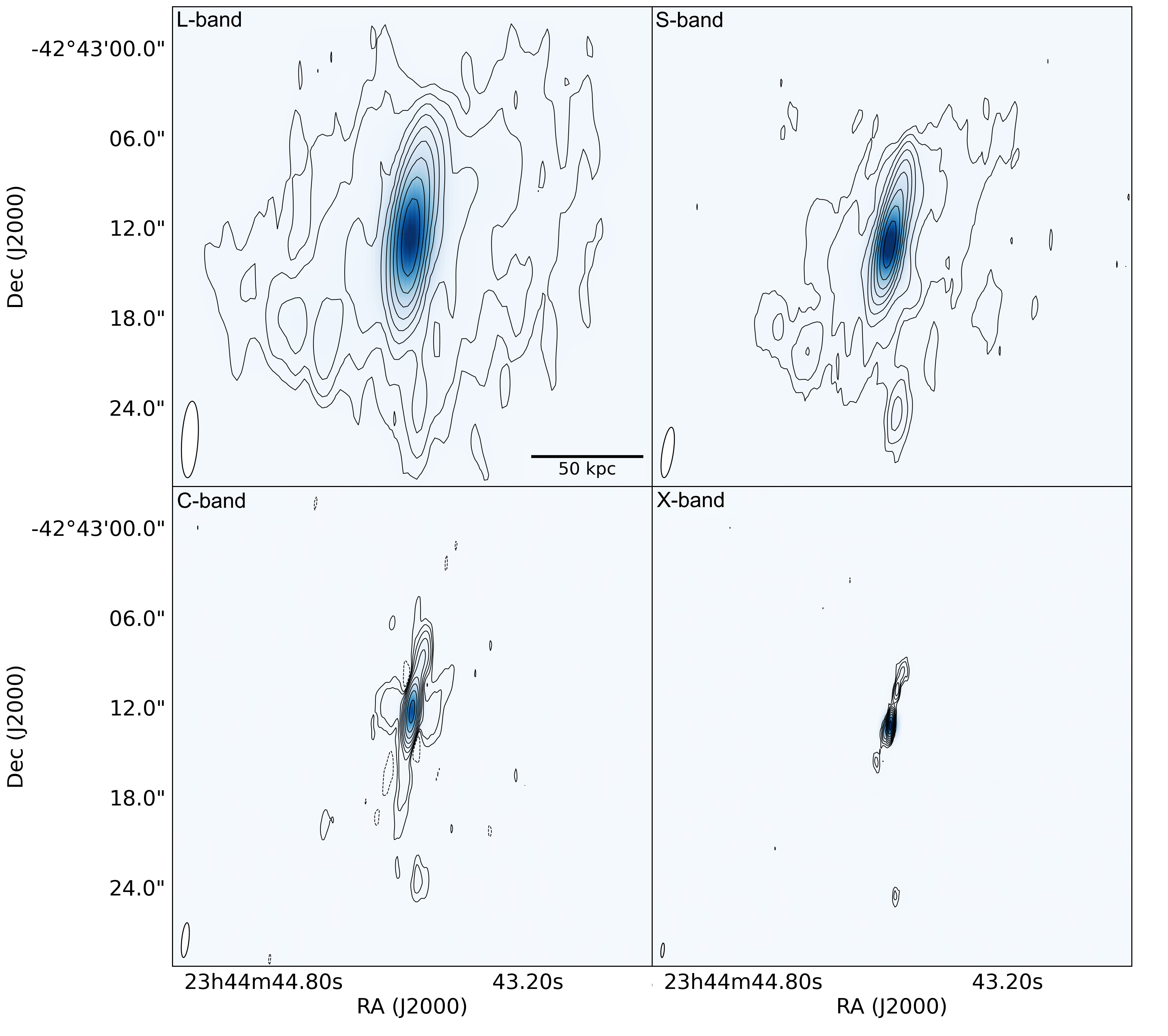}
    \caption{VLA images of the Phoenix cluster in L-band (top-left), S-band (top-right), C-band (bottom-left) and X-band (bottom-right). Contours are drawn at [-1, 1, 2, 4, 8, ...] \(\times\ 4\sigma_\textrm{rms}\), where \(\sigma_\textrm{rms}=10.8\) \textmu Jy beam\(^{-1}\) (L-band), \(\sigma_\textrm{rms}=5.9\) \textmu Jy beam\(^{-1}\) (S-band), \(\sigma_\textrm{rms}=4.3\) \textmu Jy beam\(^{-1}\) (C-band) and \(\sigma_\textrm{rms}=2.2\) \textmu Jy beam\(^{-1}\) (X-band). The beam sizes are indicated in the bottom left corners of each panel.}
    \label{fig:VLA_mosaic}
\end{figure*}

\begin{figure*}
    \centering
    \includegraphics[width=\textwidth]{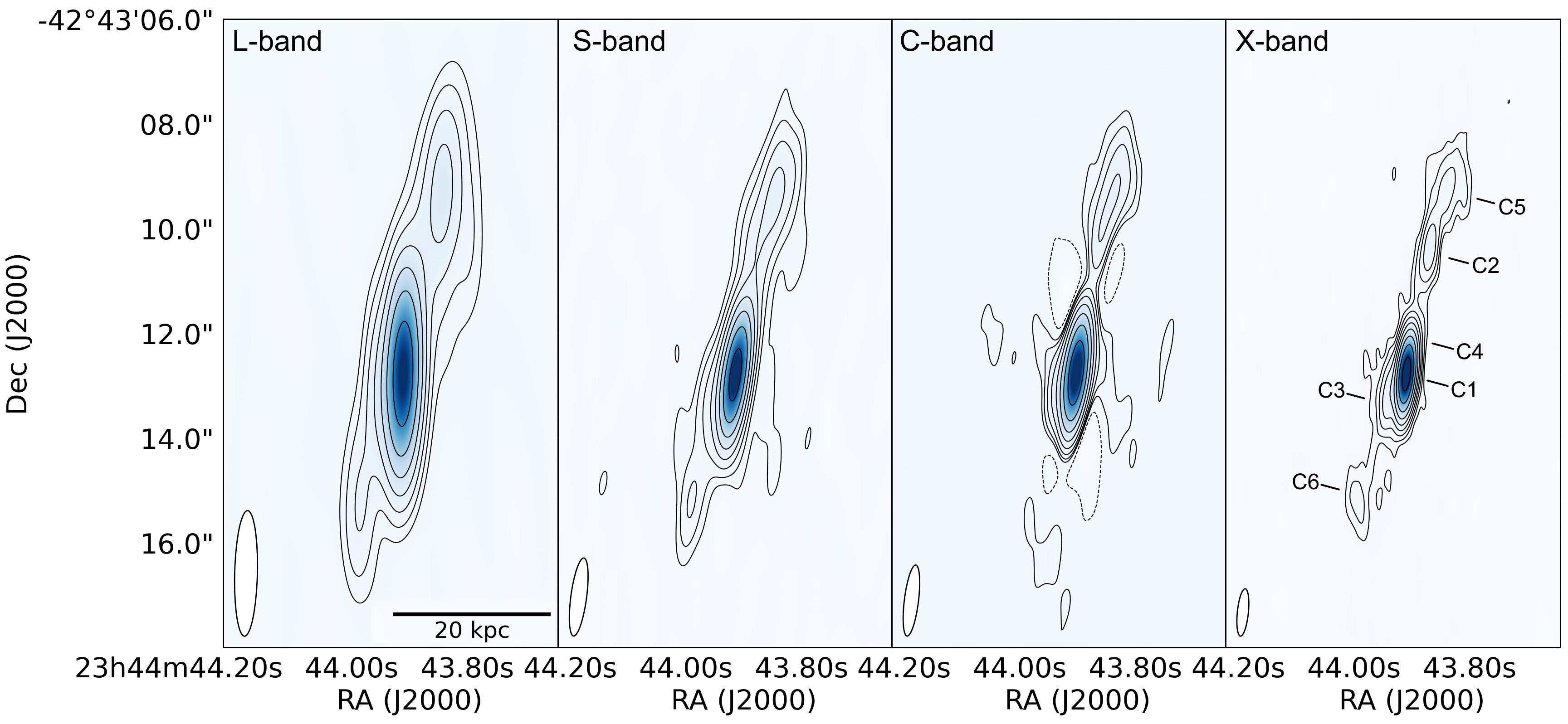}
    \caption{VLA images of the Phoenix cluster in L-band (robust -1.5), S-band (robust -1), C-band (robust -0.5) and X-band (robust 0). Contours are drawn at [-1, 1, 2, 4, 8, ...] \(\times\ 4\sigma_\textrm{rms}\), where \(\sigma_\textrm{rms}=40.8\) \textmu Jy beam\(^{-1}\) (L-band), \(\sigma_\textrm{rms}=12.8\) \textmu Jy beam\(^{-1}\) (S-band), \(\sigma_\textrm{rms}=6.9\) \textmu Jy beam\(^{-1}\) (C-band) and \(\sigma_\textrm{rms}=2.2\) \textmu Jy beam\(^{-1}\) (X-band). The source components as detected by \citet{akahori20} using the ATCA are indicated in the X-band map. The beam sizes are indicated in the bottom left corners of each panel.}
    \label{fig:VLA_mosaic_highres}
\end{figure*}

The Phoenix cluster was observed with the VLA in the L-, S-, C- and X-bands, covering the frequency range from 1 GHz to 12 GHz (PI: McDonald, 17A-258). The X-band data of this project was previously presented by \citet{mcdonald19}. In L-band, the VLA observed in both A- and B-configuration, which we complement with archival CnB-configuration observations (PI: Datta, 14B-397) recently presented by \citet{raja20}. In S-, C- and X-band, the VLA observed in A-, B- and C-configuration. The L- and S-band observations were recorded using 16 spectral windows of 64 channels each, resulting in a total bandwidth of 1 GHz and 2 GHz, respectively. The C- and X-band observations were recorded using 32 spectral windows of 64 channels each, resulting in a total bandwidth of 4 GHz for both bands. All new observations have a total of 2.5 hours of integration time per configuration. An overview of the observations is presented in Table \ref{tab:obs}.

For our new observations, we have used 3C138 and 3C147 as primary calibrators with a total integration time of approximately 5-10 minutes at the end of the observation. As the secondary calibrator, we have used J0012-3954. Scans with an integration time of approximately 2-3 minutes on the secondary calibrator were repeated every 15 minutes. In the archival L-band, CnB-array data, 3C48 was observed for 12 minutes as the primary calibrator. No secondary calibrator was included in this observation.

The data were reduced with the Common Astronomy Software Application \citep[\textsc{CASA};][]{mcmullin07}. The data reduction starts by performing Hanning smoothing on the data, flagging shadowed antennas, calculating gain elevation curves and corrections to the antenna positions, and using the \textsc{TFCrop} algorithm within \textsc{CASA} to apply automatic flagging of RFI. Next, manual flags are applied to exclude bad data from the calibration process. After the flagging, the initial complex gain solutions are calculated based on the central channels from each spectral window. These initial complex gain solutions are used to determine the delay terms. The bandpass calibration solutions are then derived based on the delay terms and the initial complex gain solutions. With the correct bandpass solutions applied, the complex gain solutions can be derived for the complete bandwidth of each spectral window. Using the polarized calibrator 3C138, we derive the global cross-hand delay solutions. The unpolarized calibrator 3C147 then allows the polarization leakage terms to be calculated. Finally, 3C138 is used again to calibrate the polarization angle. Using all relevant calibration tables, the complex gain solutions are redetermined, and the flux scale is set based on models of 3C138 and 3C147 by \citet{perleybutler12}. The calibration solutions are then applied to the target source, after which the \textsc{TFCrop} and \textsc{RFlag} automatic flagging algorithms are used to remove previously undetected RFI. Next, the calibrated data of the target source are split out, after which the \textsc{AOFlagger} software \citep{offringa10} is used to remove any remaining RFI.

Finally, we improve the calibration through the process of self-calibration. We use \textsc{CASA} for calculating new calibration solutions and applying these to the data, and we use \textsc{WSClean} \citep{offringa14} for the imaging and deconvolution. Each data set has been self-calibrated by initially performing phase-only self-calibration, and later performing amplitude \& phase self-calibration, with iteratively shorter calibration solutions. Then, all data sets of the same spectral band are concatenated to form one data set per band. These data set are then self-calibrated again, to obtain the final data sets. Imaging is performed using Briggs weighting \citep{briggs95} with a robust parameter of zero. The C-band data experienced issues during calibration, which is suspected to be caused by the very low declination of the source. For this reason, it is difficult to discern real structures from noise features near the AGN.

\section{Results}

The images we obtain are shown in Figure \ref{fig:VLA_mosaic}. The L-band image shows the central compact AGN with the diffuse mini-halo surrounding it. The mini-halo is less visible in the S-band image due to the spectral slope of the mini-halo, as well as the more compact beam. In the C-band image, the jetted outflows are visible towards the North and the South of the AGN. A small part of the mini-halo is still visible towards the East and West of the AGN. Finally, the X-band image shows the jets at the highest angular resolution. The total flux densities, peak fluxes, rms noise levels and beam sizes of the final images are summarized in Table \ref{tab:obs_final}.

To show the maximum resolution attainable with each of the four data sets, images of the target using lower robust parameters are shown in Figure \ref{fig:VLA_mosaic_highres}. This shows that using a different weighting scheme, the jetted outflows can even be resolved in L-band. Comparing our observations to the ATCA observations of \citet{akahori20}, we find that our VLA observations are able to resolve all components observed with the ATCA: C1 (AGN core), C2, C4 and C5 (northern lobe) and C3 and C6 (southern lobe). \citet{akahori20} mention that they possibly detect jet precession, as components C3 and C4 (near the AGN), appear to be emitted in a different direction than components C2 and C6 in the northern and southern jets, respectively.

\begin{table*}
    \centering
    \caption{Properties of the images shown in Figure \ref{fig:VLA_mosaic}}
    \begin{tabular}{lllllll}
        \hline\hline
        Freq. band & Flux density & Peak flux & rms noise & b\(_\text{major}\) & b\(_\text{minor}\) & b\(_\text{PA}\)\\
         & (mJy) & (mJy beam\(^{-1}\)) & (\textmu Jy beam\(^{-1}\)) & (") & (") & (deg)\\\hline
        L-band & 33.8 & 20.4 & 10.8 & 5.1 & 1.1 & -3.7\\
        S-band & 18.0 & 10.5 & 5.9 & 3.4 & 0.74 & -8.6\\
        C-band & 7.95 & 6.46 & 4.3 & 2.5 & 0.51 & -5.9\\
        X-band & 4.01 & 3.20 & 2.2 & 0.96 & 0.22 & -5.6\\\hline\hline
    \end{tabular}
    \label{tab:obs_final}
\end{table*}

\begin{figure}
    \centering
    \includegraphics[width=\columnwidth]{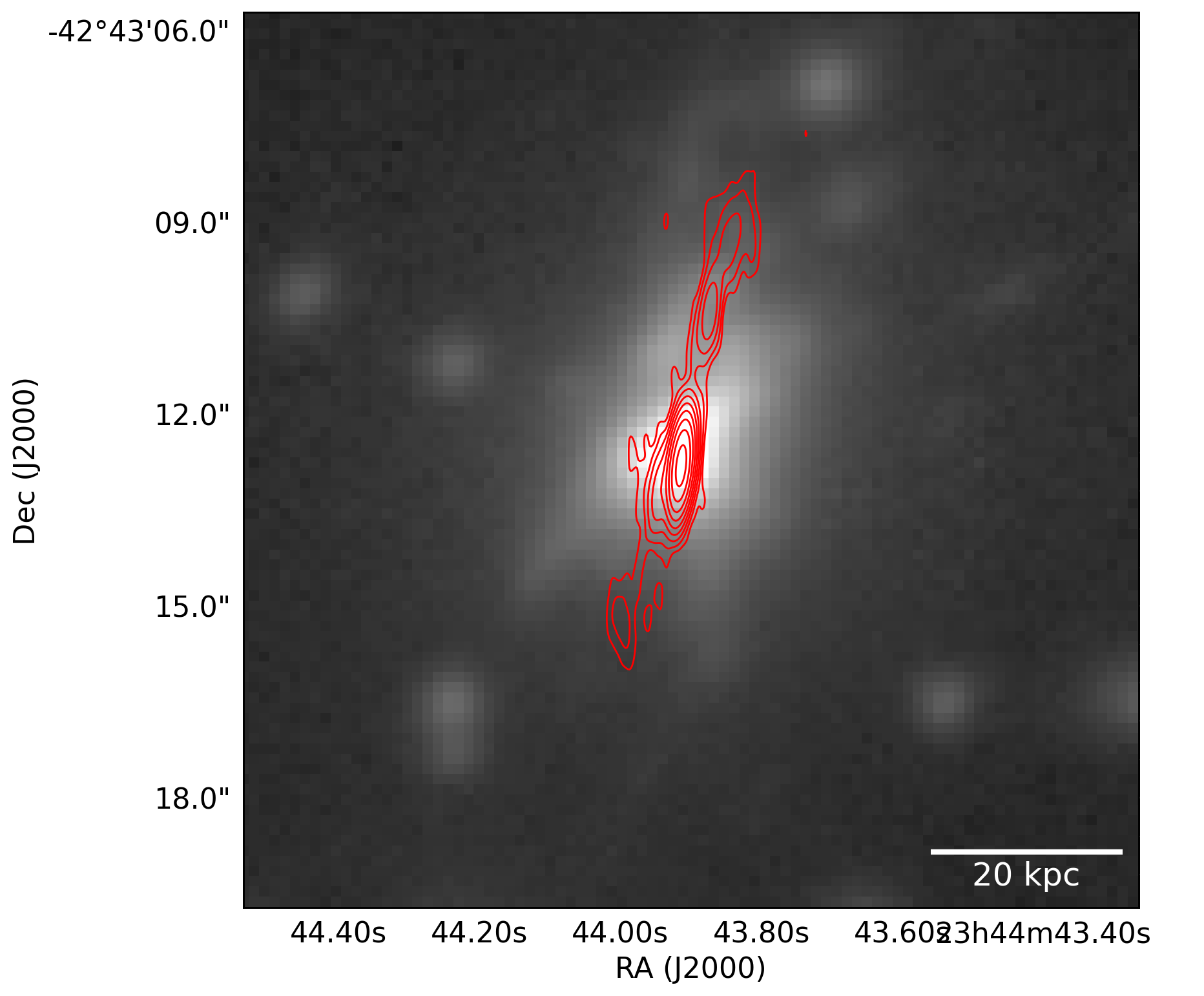}
    \caption{Optical r-band image of the Phoenix cluster taken with Megacam on the Magellan Clay Telescope \citep{mcdonald15}. The red contours indicate the X-band emission, and are drawn at [-1, 1, 2, 4, 8, ...] \(\times\ 4\sigma_\textrm{rms}\), where \(\sigma_\textrm{rms}=2.2\) \textmu Jy beam\(^{-1}\)}
    \label{fig:optical_overlay}
\end{figure}

\begin{figure}
    \centering
    \includegraphics[width=\columnwidth]{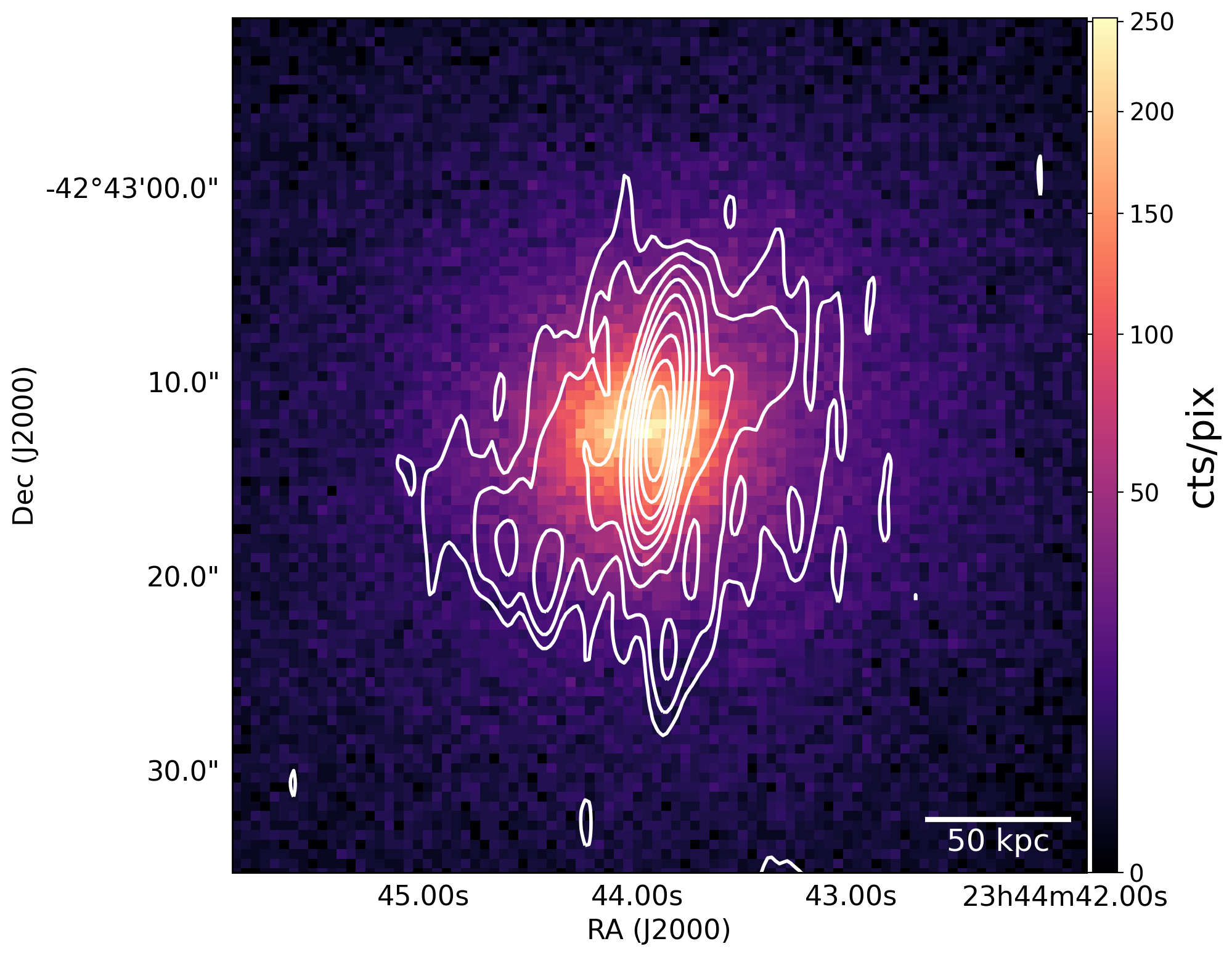}
    \caption{X-ray (0.7-2 keV) image by \textit{Chandra} \citep{mcdonald15}. The white contours indicate the L-band emission as observed with the \textit{Very Large Array}, and are drawn at [-1, 1, 2, 4, 8, ...] \(\times\ 4\sigma_\textrm{rms}\), where \(\sigma_\textrm{rms}=10.8\) \textmu Jy beam\(^{-1}\)}
    \label{fig:xray_lband}
\end{figure}

\begin{figure}
    \centering
    \includegraphics[width=\columnwidth]{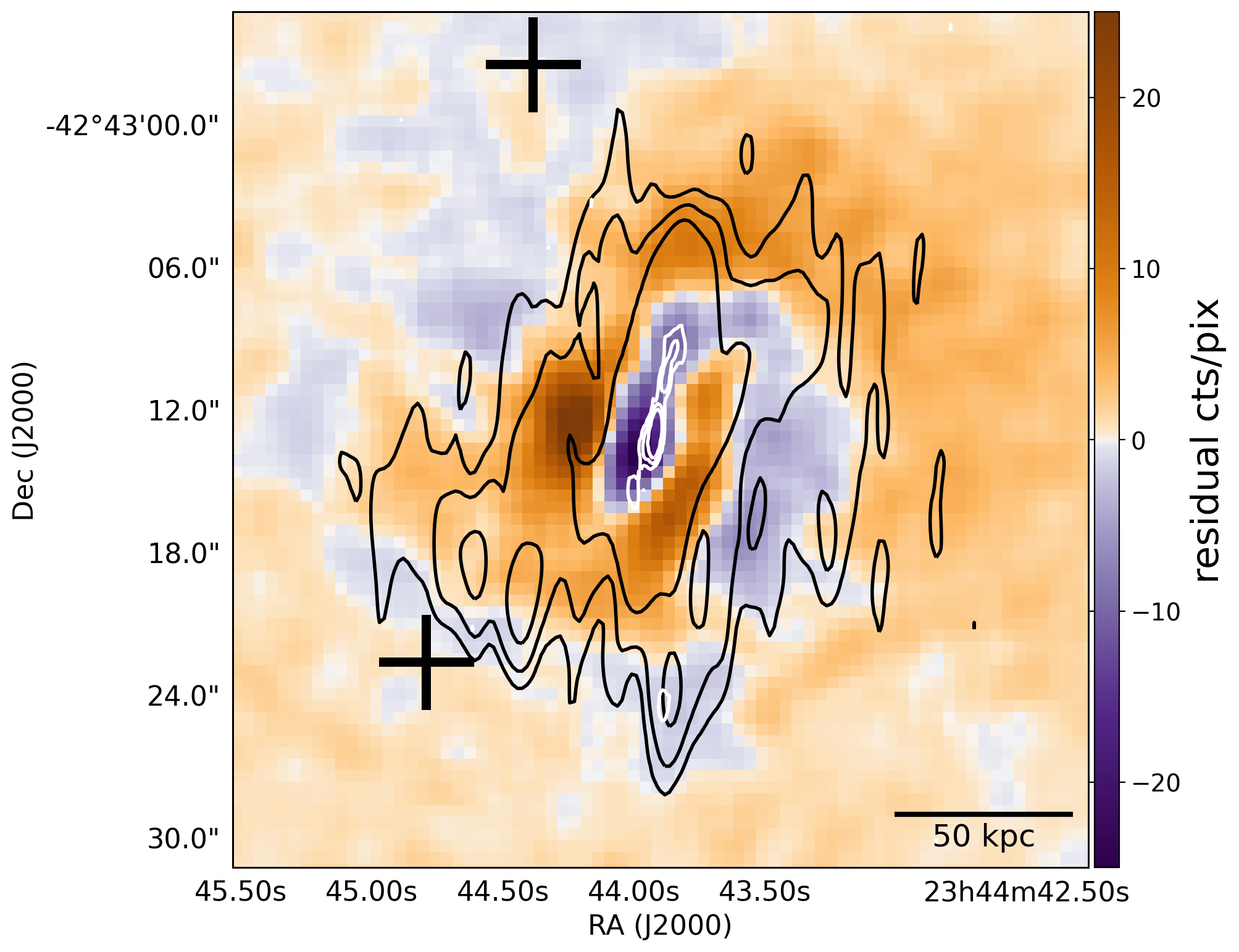}
    \caption{Residuals of the \textit{Chandra} X-ray image \citep{mcdonald15} minus a \(\beta\)-model. The residuals have been smoothed by a boxcar function with a scale of 3 pixels (1.5 arcsec). The white contours indicate the emission in X-band as observed with the VLA, and are drawn at [-1, 1, 4, 16] \(\times\ 4\sigma_\textrm{rms}\), where \(\sigma_\textrm{rms}=2.2\) \textmu Jy beam\(^{-1}\). The black contours indicate the emission in L-band as observed with the VLA, and are drawn at [-1, 1, 2, 4] \(\times\ 4\sigma_\textrm{rms}\), where \(\sigma_\textrm{rms}=10.8\) \textmu Jy beam\(^{-1}\). The black plusses indicate the positions of the ghost cavities detected by \citet{mcdonald15}.}
    \label{fig:Xray_res_LXband_overlay}
\end{figure}

The emission from the AGN appears to be coincident with the BCG, as shown in Figure \ref{fig:optical_overlay}. In addition, the L-band emission extends far beyond the optical size of the BCG and the star-forming filaments. The radio emission from the cluster coincides with X-ray emission detected by \textit{Chandra} \citep{mcdonald15}, as shown in Figure \ref{fig:xray_lband}. To confirm that the jetted outflows detected in our VLA observations are coincident with the cavities previously detected in the ICM, we subtract a \(\beta\)-model from the X-ray map, and then overlay the L- and X-band contours on the residuals, as shown in Figure \ref{fig:Xray_res_LXband_overlay}. Although there is a small sub-arcsecond uncertainty in the relative alignment between the images, it is clear that the X-ray cavities are inflated by magnetized radio plasma. We find no radio emission coincident with the possible ghost cavities previously marginally detected by \citet{mcdonald15} further towards the north-east and south-east.

\begin{figure}
    \centering
    \includegraphics[width=\columnwidth]{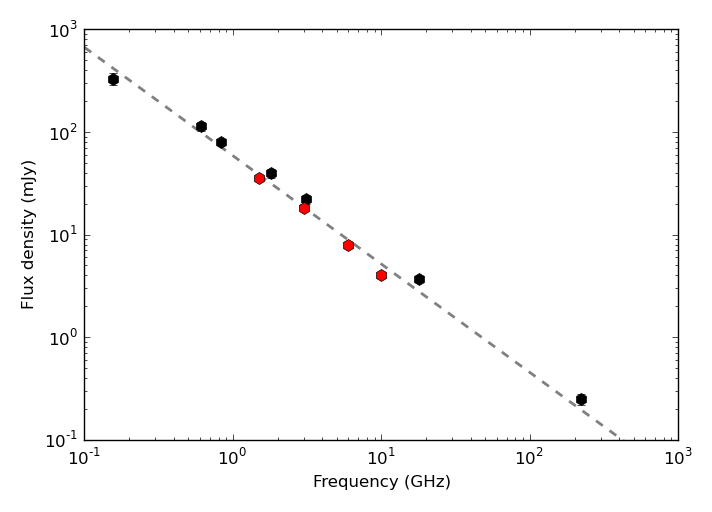}
    \caption{Spectral energy distribution of the Phoenix cluster. The black dots show data obtained from literature. The red dots show data presented in this work. The best-fitting spectral index through the data is \(-1.12 \pm 0.02\). A curvature term is included in the fit, but is found to be negligible. The data point at 220 GHz is excluded from the fit as free-free emission and thermal dust emission are expected to contribute significantly at this frequency.}
    \label{fig:sed}
\end{figure}

To derive the overall spectral index of the source, we combine measurements of the total flux density of the cluster in our four bands with archival data from \citet{mauch03, mcdonald14, vanweeren14, akahori20}. We assume a 5\% uncertainty on our flux density estimates in accordance with \citet{perleybutler17}. By fitting a power-law profile through the data, we obtain an overall spectral index of \(-1.12\ \pm\ 0.02\), as shown in Figure \ref{fig:sed}. To account for a possible curvature in the spectrum, we fit a second-degree polynomial in log-space through the data, but find the curvature term to be consistent with zero within the 95\% confidence interval. We exclude the data point at 220 GHz from this fit, as we expect that free-free emission and thermal dust emission can significantly contribute to the spectrum at such high frequencies, causing the model of a single power-law to break down.

To study the mini-halo and the AGN, we need to separate these two components. By producing a map of the source using only the long baselines, we obtain an image of the AGN and its jets, without contamination from the more extended mini-halo. Based on the X-band imaging, we find that the AGN and its jets are more compact than an angular scale of 10~arcseconds, which corresponds to a limit on the baseline length of 20~k\(\lambda\). After producing an L-band image using only baselines longer than 20 k\(\lambda\), we find that the remaining flux density of the source is 25.2~mJy within the \(3\sigma\) contours. Comparing this flux density to the total flux density of 33.8~mJy, we find that there is a discrepancy of 8.6~mJy, which we attribute to the mini-halo.

Alternatively, the flux density of the mini-halo can be estimated through radial profile fitting by describing both the AGN and the mini-halo using circular Gaussians. This enables the flux of the AGN and the mini-halo to be spatially disentangled, thereby potentially providing a more accurate measurement of their total flux densities. We produce an L-band image with a Briggs robust parameter of -1 to improve the resolution of the image, and thereby reduce the scale of the central AGN. This image is smoothed to a circular beam of 2.8~arcseconds based on the major axis of the image. Through least-squares fitting, we obtain separate estimates for the flux density of the AGN and the mini-halo. For the AGN, we find a flux density of \(23.6\ \pm\ 2.4\)~mJy, whereas we find a flux density of \(8.5\ \pm\ 0.9\)~mJy for the mini-halo.

Similarly, we can perform a radial fit with all compact emission masked out, as we know that the jetted outflows are only present in the Northern and Southern directions, which can cause a systematic error. By using the L-band map produced using only long baselines, we can define a mask of where we expect the source to be dominated by compact structure. Outside of this mask, the mini-halo is expected to be the dominant component. We define this mask as the 3\(\sigma\) region in the 20 k\(\lambda\) L-band map. Using this mask, we can fit the source using a single Gaussian profile to represent the mini-halo. The estimate obtained with a mask for the flux density of the mini-halo is considerably lower, at only 5.9~mJy. As the values obtained from the first two methods agree very well, we adopt a value of 8.5~mJy for the flux density of the mini-halo in L-band for the rest of this paper.

From the radial fitting process, we can also obtain an estimate for the extent of the mini-halo. From the radial fit, we find that the mini-halo can be described by a Gaussian with a FWHM of 13.8~arcseconds, deconvolved with the beam. At a redshift of \(z=0.597\), this corresponds to a scale of 92~kpc. The maximum observable radius of the mini-halo - defined as the radius at which the mini-halo reaches the noise level - is 17.8~arseconds, or about 120~kpc. This gives a total diameter of the mini-halo of \(\sim\)240~kpc.

\begin{figure*}
    \centering
    \includegraphics[width=\textwidth]{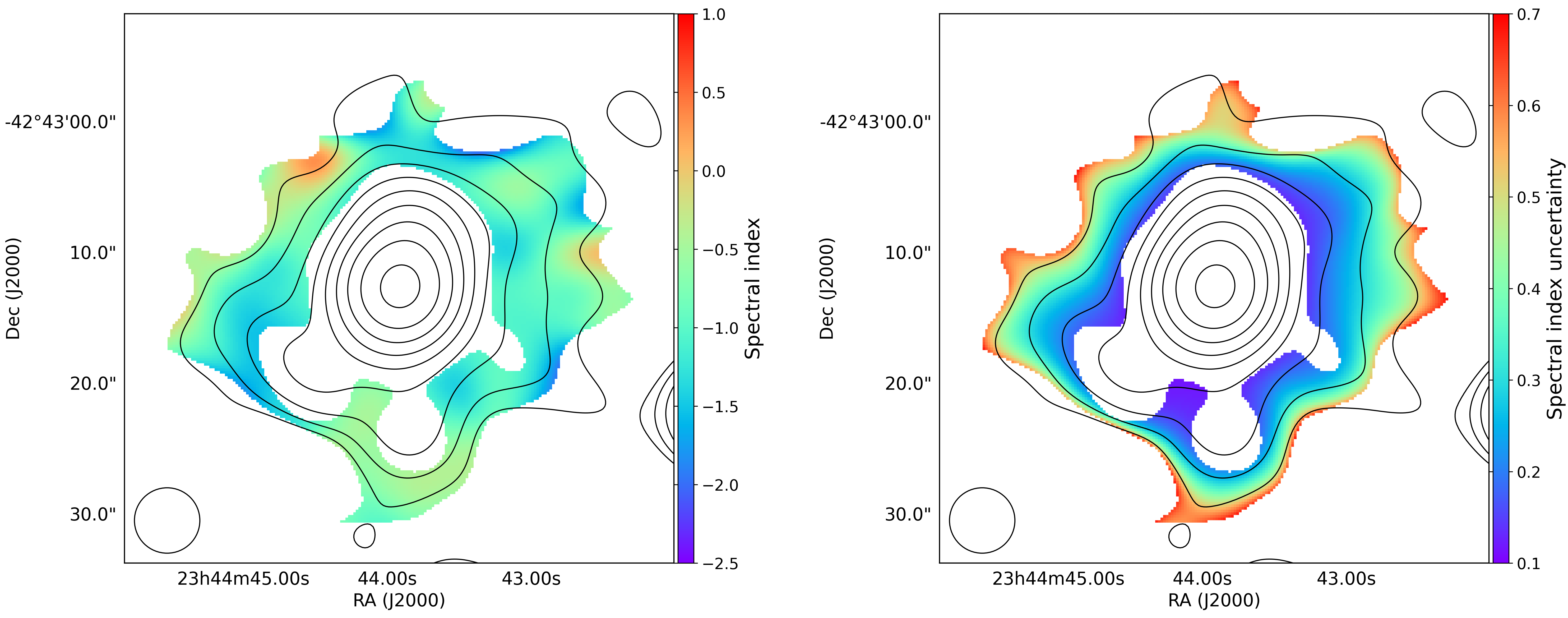}
    \caption{\textit{Left:} spectral index map from the L- and S-band images. \textit{Right:} corresponding uncertainties of the spectral index map. The contours show the L-band image smoothed to a resolution of 5 arcseconds and are drawn at [-1, 1, 2, 4, 8, ...] \(\times\ 4\sigma_\textrm{rms}\), where \(\sigma_\textrm{rms}=17.9\) \textmu Jy beam\(^{-1}\). The circular beam of 5 arcseconds is shown in the bottom-left corner.}
    \label{fig:LS_spixmap}
\end{figure*}

To map the spectral index of the halo, we smooth both the L-band and the S-band images to a circular beam size of 5 arcseconds and align the images by matching the positions of a nearby point source. To avoid contamination from the AGN and its jets, we mask out the central region using the same mask as with the radial fitting. After calculating the spectral index between the two maps and excluding the masked region, we obtain the spectral index map of the mini-halo as shown in Figure \ref{fig:LS_spixmap}.

The spectral index map of the mini-halo shows an annulus with a mean spectral index of \(\alpha\ =\ -0.95 \pm 0.10\). However, the signal-to-noise ratio is too low to provide insight about any potential gradients or cut-offs in the spectrum as a function of radius. Using this spectral index, we calculate the radio luminosity of the mini-halo using

\begin{equation}\label{eq:radiolum}\centering
    P_\mathrm{1.4 GHz}\ =\ 4\pi S_\mathrm{1.4 GHz} D_L^2(1+z)^{-\alpha-1},
\end{equation}

\noindent where \(S_\mathrm{1.4 GHz}\) is the flux density at 1.4 GHz and \(D_L\) is the luminosity distance to the source, and find a value of \(P_\mathrm{1.4 GHz}=(13.0 \pm 1.4)\times 10^{24}\ \mathrm{W}\ \mathrm{Hz}^{-1}\).

\begin{figure*}
    \centering
    \includegraphics[width=\textwidth]{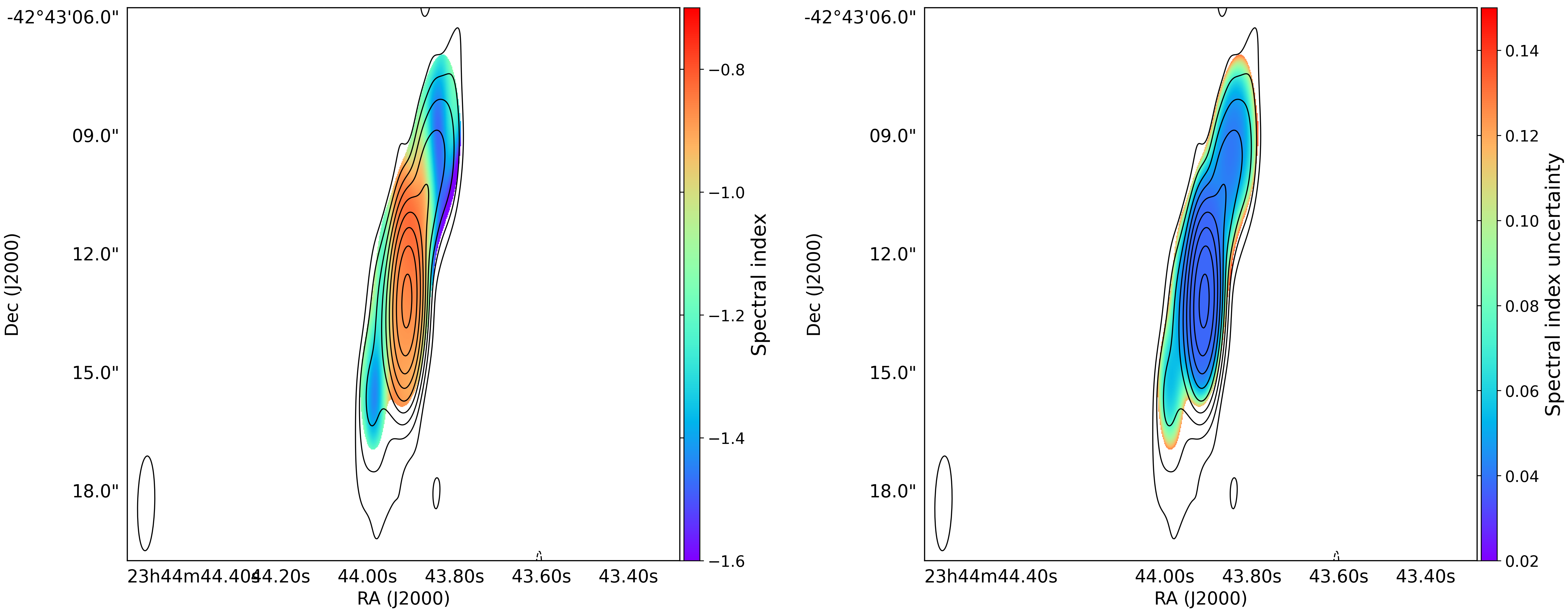}
    \caption{\textit{Left:} Spectral index map from the L- and X-band images. \textit{Right:} corresponding uncertainties of the spectral index map. The contours show the X-band image smoothed to the L-band resolution and are drawn at [-1, 1, 2, 4, 8, ...] \(\times\ 4\sigma_\textrm{rms}\), where \(\sigma_\textrm{rms}=2.6\) \textmu Jy beam\(^{-1}\). The beam is shown in the bottom-left corner.}
    \label{fig:LX_spixmap}
\end{figure*}

To map the spectral index of the AGN and its lobes, we take the high-resolution L-band image, as shown in Figure \ref{fig:VLA_mosaic_highres}, and smooth the X-band image to this resolution. Next, we align the images based on a nearby point source. By calculating the spectral index between L-band and X-band, we obtain the map shown in Figure \ref{fig:LX_spixmap}.

In this spectral index map, we can resolve both the spectral indices of the two lobes, as well as the spectral index of the AGN. We find that the northern lobe has a spectral index between L- and X-band of -1.35 \(\pm\) 0.07, and the southern lobe has a spectral index of -1.30 \(\pm\) 0.12. The AGN in the center of the map shows a spectral index of -0.86.\(\pm\) 0.04. Due to the alignment of the lobes with the synthesized beam, we do not have the resolution required to check for a spectral gradient along the outflows. By subtracting a point source convolved with the beam from the high-resolution L-band image, we find that the radio lobes have a total flux density of \(7.6 \pm 0.8\)~mJy.

Finally, as we have VLA observations in full polarization mode, we have checked for polarized emission using RM synthesis, but were unable found a significant amount of polarized emission from the cluster, consistent with the results of \citet{akahori20}.

\begin{figure*}
    \centering
    \includegraphics[width=\textwidth]{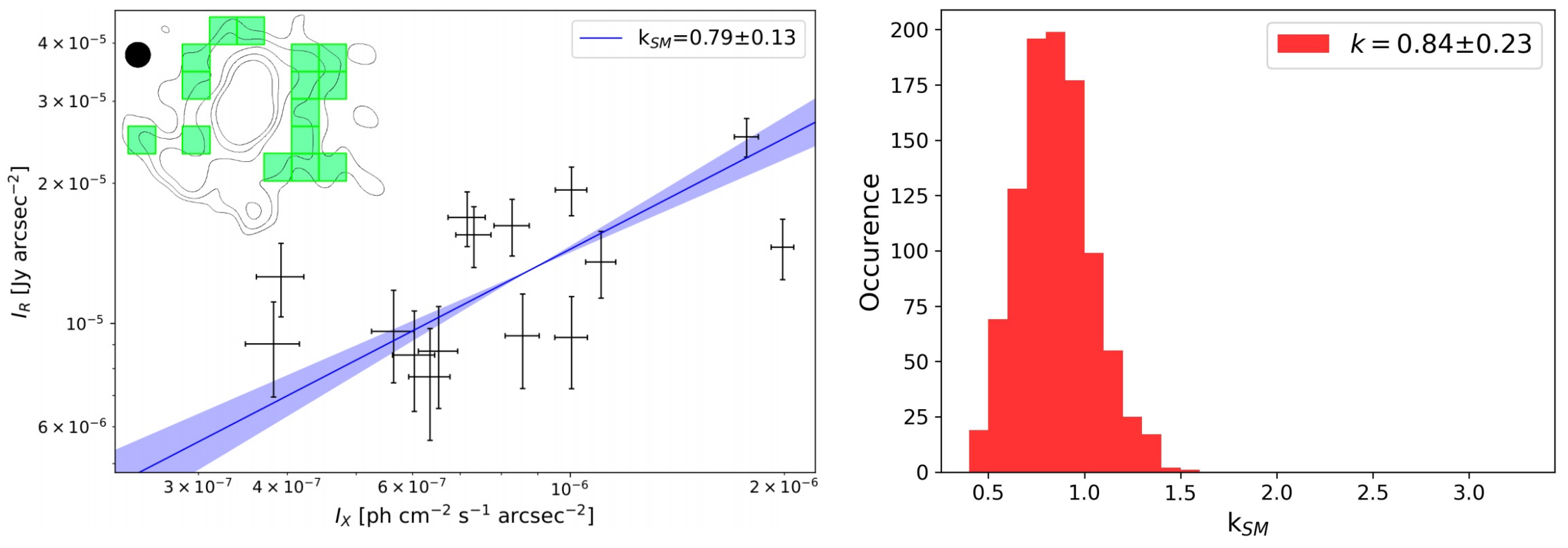}
    \caption{Results from the Monte Carlo point-to-point analysis. \textit{Left:} an example of a estimate for \(k\) using a particular choice for the grid position. The data points and error bars show the estimated radio surface brightness vs. X-ray surface brightness for a given grid cell. The slope of the best-fitting power-law through these data (blue line) is shown in the legend. The contours in the top-left corner indicate the [3,6,12,24,48] \(\times \sigma\) contours of the circularly-smoothed L-band image with a Briggs robust parameter of -1, where \(\sigma=18\) \textmu Jy. The beam size is 2.8 arcseconds circular, and is indicated by the solid black circle. \textit{Right:} histogram of all values of \(k\) from the Monte Carlo point-to-point analysis. The resulting best estimate for \(k_\mathrm{MC}\) is reported in the legend.}
    \label{fig:p2p}
\end{figure*}

\section{Discussion}

\subsection{The origin of the mini-halo}

Our VLA observations clearly resolve the mini-halo and the AGN in the Phoenix cluster. We estimate the mini-halo to have a maximum observable deconvolved diameter of about 240 kpc, and a radio luminosity at 1.4 GHz of \(P_\mathrm{1.4 GHz}=(13.0 \pm 1.4)\times 10^{24}\ \mathrm{W}\ \mathrm{Hz}^{-1}\). Our estimate for the size of the mini-halo is smaller than the estimate by \citet{vanweeren14}, who estimate a size in the range of 400-500 kpc using 610 MHz GMRT observations. This may be an indication of spectral steepening in the outer regions of the mini-halo, as a detection by the GMRT at 610 MHz and a non-detection by the VLA at 1.5 GHz in this region implies a spectral index steeper than \(\alpha=-1.5\), based on the rms noise levels in both maps. However, we do not see a trend in our spectral index maps that suggests such a spectral steepening. In addition, the GMRT data suffers from a relatively poor angular resolution and sensitivity compared to the VLA, so we can not make any definite claims on the spectrum of the mini-halo in the outer regions.

Our reported value for the radio luminosity at 1.4 GHz is consistent with previous estimates by \citet{vanweeren14} and \citet{raja20}, who calculate values of \(P_\mathrm{1.4 GHz}=(10.4\pm3.5)\times 10^{24} \mathrm{W} \mathrm{Hz}^{-1}\) and \(P_\mathrm{1.4 GHz}=(14.38\pm1.80)\times 10^{24} \mathrm{W} \mathrm{Hz}^{-1}\), respectively.

We have mapped the spectral index of the mini-halo, and derive an integrated spectral index between L- and S-band of -0.95 \(\pm\) 0.10, which is consistent with the spectral index of -0.98 \(\pm\) 0.16 as derived by \citet{raja20} between 610 MHz and 1.5 GHz. We were unable to find evidence for a radial gradient in the spectral index map due to the low signal-to-noise ratio. 

Further insight into the origin of the diffuse emission can be provided by the spatial correlation between the radio (\(I_R\)) and X-ray (\(I_X\)) surface brightnesses. This correlation is expected because the relativistic electron population, and hence the radio emission, is predicted to be linked to the thermal plasma in both hadronic and re-acceleration models. The \(I_R\)-\(I_X\) correlation allows us to constrain the distribution of the non-thermal ICM components with respect to the thermal plasma and thereby investigate the origin of the radio emission \citep[e.g.][]{brunetti14, ignesti20}.

We have used the Monte Carlo point-to-point analysis presented in \citet{ignesti20} to evaluate the \(I_R\)-\(I_X\) correlation for the Phoenix cluster. We use a circularly-smoothed image of the Phoenix cluster in L-band with a Briggs robust parameter of -1 to improve the resolution of the map, and thereby reduce the area affected by AGN-related emission while simultaneously increasing the amount of samples of mini-halo emission. The X-ray surface brightness is obtained from archival Chandra observations. The surface brightnesses \(I_R\) and \(I_X\) have been sampled with 1000 randomly-generated meshes. For the cells in the mini-halo region, values of \(I_R\) and \(I_X\) are measured and fitted with a power-law relation \(I_R\propto I_X^k\) using the BCES algorithm \citep{akritas96}. We present in Figure \ref{fig:p2p} both the result of the analysis performed on a single grid (left panel) and the final result of the MC routine (right panel). We measure a sub-linear scaling index \(k\) of \(0.84 \pm 0.23\), which indicates that the radio emission likely declines slower than the X-ray emission. This result is interestingly more similar to what is observed for giant radio halos, which often feature sub-linear indices \(k\) in the range of 0.5 to 1.0 \citep{govoni01, feretti01, giacintucci05, hoang19, xie20}, than to mini-halos, which generally feature a super-linear scaling with indices \(k\) in the range of 1.1 to 1.3 \citep{ignesti20}.
Such a flat index \(k\) indicates that the X-ray emission is more peaked than the radio emission, which is in agreement with the exceptionally luminous cool core of this cluster.  Therefore, it suggests that the distribution of non-thermal components does not strongly depend on the properties of the cool core.

\begin{figure}
    \centering
    \includegraphics[width=\columnwidth]{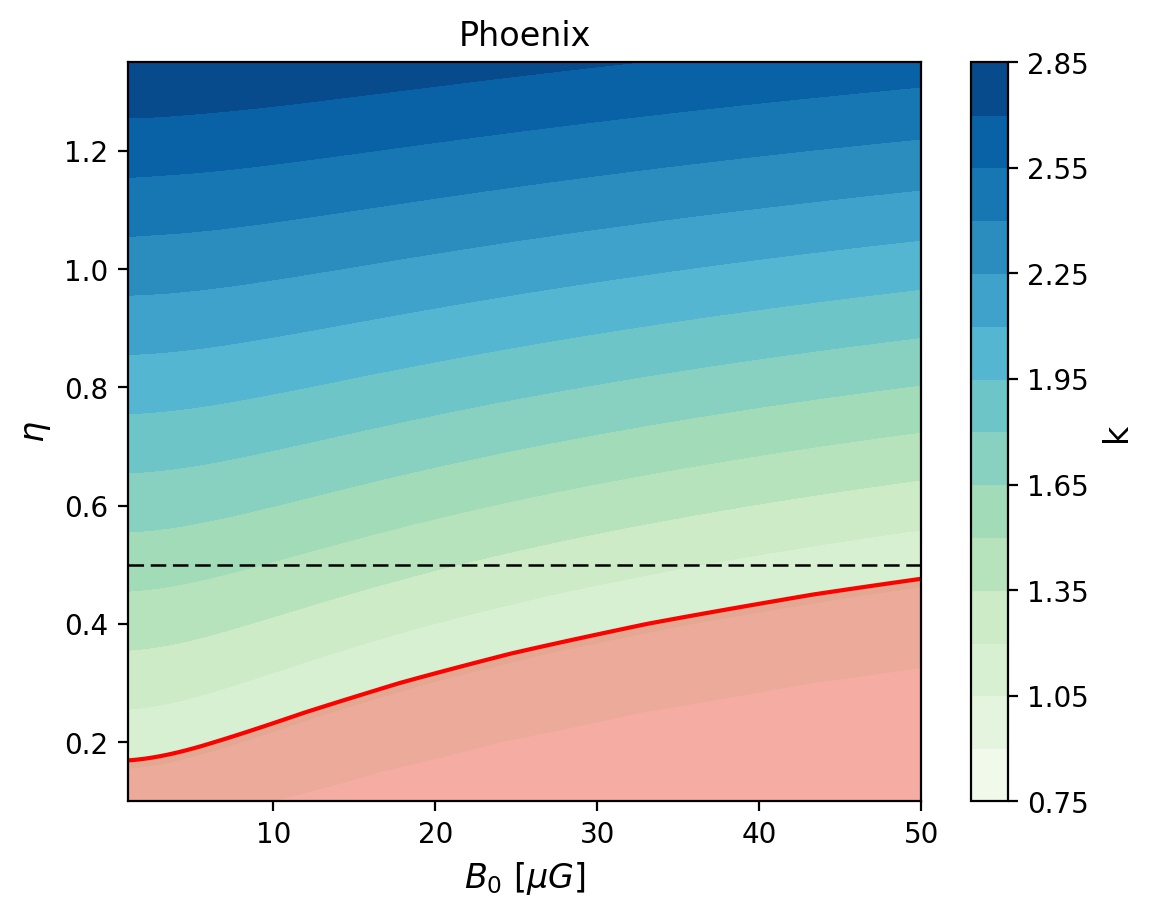}
    \caption{Index \(k\) for combinations of the magnetic field strength and the index \(\eta\). The horizontal dashed line at \(\eta\)=0.5 indicates the equilibrium configuration between thermal and non-thermal energy density. The red region indicates the parameter space consistent with values of \(k\) we obtain using the Monte Carlo point-to-point analysis.}
    \label{fig:eta_k}
\end{figure}

On the basis of these results, we can explore the scenario of purely hadronic origin of relativistic electrons. We follow the approach presented in \citet{ignesti20} to infer the ICM magnetic field to constrain the physical boundaries of the hadronic model for the Phoenix cluster. We use the thermodynamic profiles from \citet{mcdonald15}, to compute the ICM electron density and temperature. From these, we numerically calculate the X-ray emissivity. Assuming a hadronic model, we can compare these to the radio emissivity expected in a pure hadronic model \citep[e.g.][]{brunetti12} to constrain the magnetic field configurations that are consistent with our value of the index \(k\). We assume a magnetic field profile of the form

\begin{equation}\label{eq:mag}
    B(r) = B_0\left(\frac{n(r)}{n_0}\right)^\eta,
\end{equation}

\noindent where \(B_0\) and \(n_0\) are the central values of the magnetic field strength and the thermal ICM number density, respectively, and the index \(\eta\) determines the scaling relation between the magnetic field and the ICM density. The constraints we derive on the magnetic field configuration are shown in Figure \ref{fig:eta_k}. We find that for typical values of the central magnetic field strength \citep[10-20 \textmu G;][]{carilli02}, we require the index \(\eta\) to be 0.2-0.3, indicating that the ICM density is much more peaked than the magnetic field strength. Similarly, for a typical value of \(\eta=0.5\), we require the central magnetic field strength to be at least 50 \textmu G, which is far higher than commonly observed.

To provide context to these estimates of the magnetic field configuration from the point-to-point analysis, we also calculate the equipartition magnetic field. Under the assumption that the magnetic field and the cosmic ray particles evolve on similar timescales, they are expected to be coupled. This coupling is supposed to lead to an equilibrium between the energy densities of the cosmic ray particles and the magnetic field \citep{govoni04, beck05}. For the classical equipartition magnetic field, we obtain a field strength of 3.4 \textmu Gauss. For the revised equipartition magnetic field, we obtain a field strength of 5.5 \textmu Gauss using a commonly adopted value of \(\gamma_\mathrm{min}=100\) for the low-energy cutoff of the cosmic ray particle energy distribution. This shows that the equipartition magnetic field strength is much lower than the magnetic field strength predicted by a hadronic model.

\begin{figure}
    \centering
    \includegraphics[width=\columnwidth]{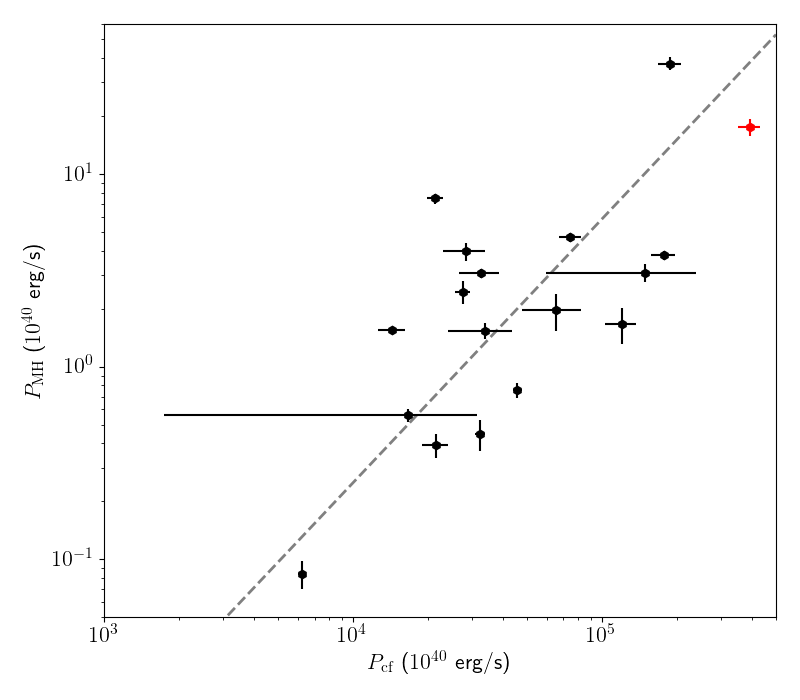}
    \caption{Cooling flow power versus mini-halo radio luminosity at 1.4 GHz for the sample of mini-halos from \citet{giacintucci19}. The red data point indicates the Phoenix cluster. The dashed grey line indicates the best power-law fit through the data, and is given by \(\log P_\text{MH} [10^{40} \text{erg/s}] = (1.37 \pm 0.17)\log P_\text{cf} [10^{40} \text{erg/s}] - (6.07 \pm 0.79)\). The integrated mini-halo luminosities are from \citet{giacintucci19} and this work. Values of the cooling rates and ICM temperatures are from \citet{arnaud87}, \citet{churazov03}, \citet{gitti04b}, \citet{bohringer05}, \citet{covone06}, \citet{leccardi08}, \citet{bravi15}, \citet{mcdonald15}, \citet{main16}, \citet{werner16} and \citet{mcdonald18}}
    \label{fig:Pcf_PMH}
\end{figure}

\begin{figure}
    \centering
    \includegraphics[width=\columnwidth]{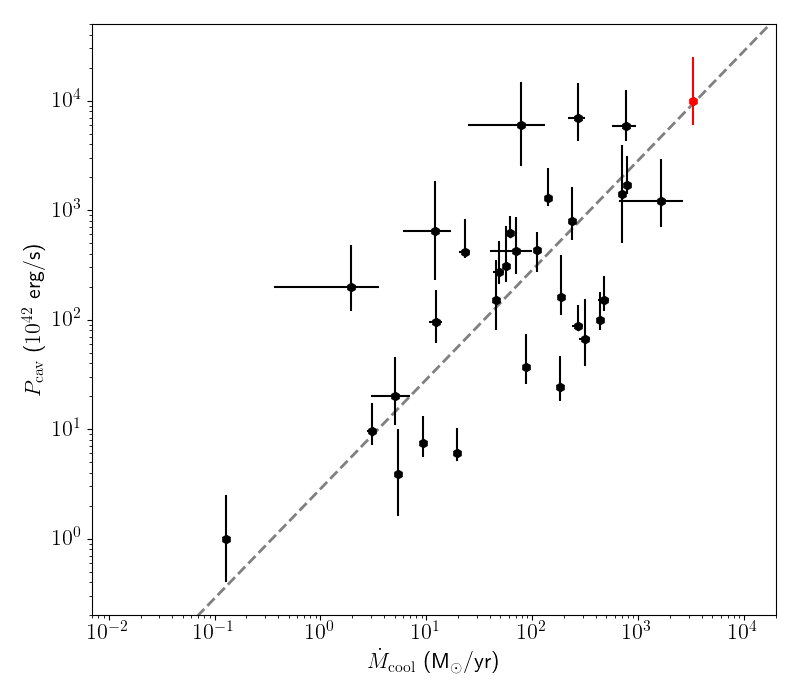}
    \caption{ICM cooling rate \(\dot{M}_\mathrm{cool}\) versus cavity power \(P_\mathrm{cav}\) for the sample of clusters from \citet{rafferty06}. The red data point indicates the Phoenix cluster. The grey dashed line indicates the best power-law fit through the data, and is given by \(\log P_\text{cav} [10^{42} \text{erg/s}] = (1.00 \pm 0.10)\log \dot{M}_\text{cool} [\text{M}_\odot / \text{yr}] + (0.45 \pm 0.24)\). All black data points are from \citet{rafferty06} and \citet{mcdonald18}, and the data for the Phoenix cluster are from \citet{mcdonald15, mcdonald19}.}
    \label{fig:Mdot_Pcav}
\end{figure}

\begin{figure*}
    \centering
    \includegraphics[width=\textwidth]{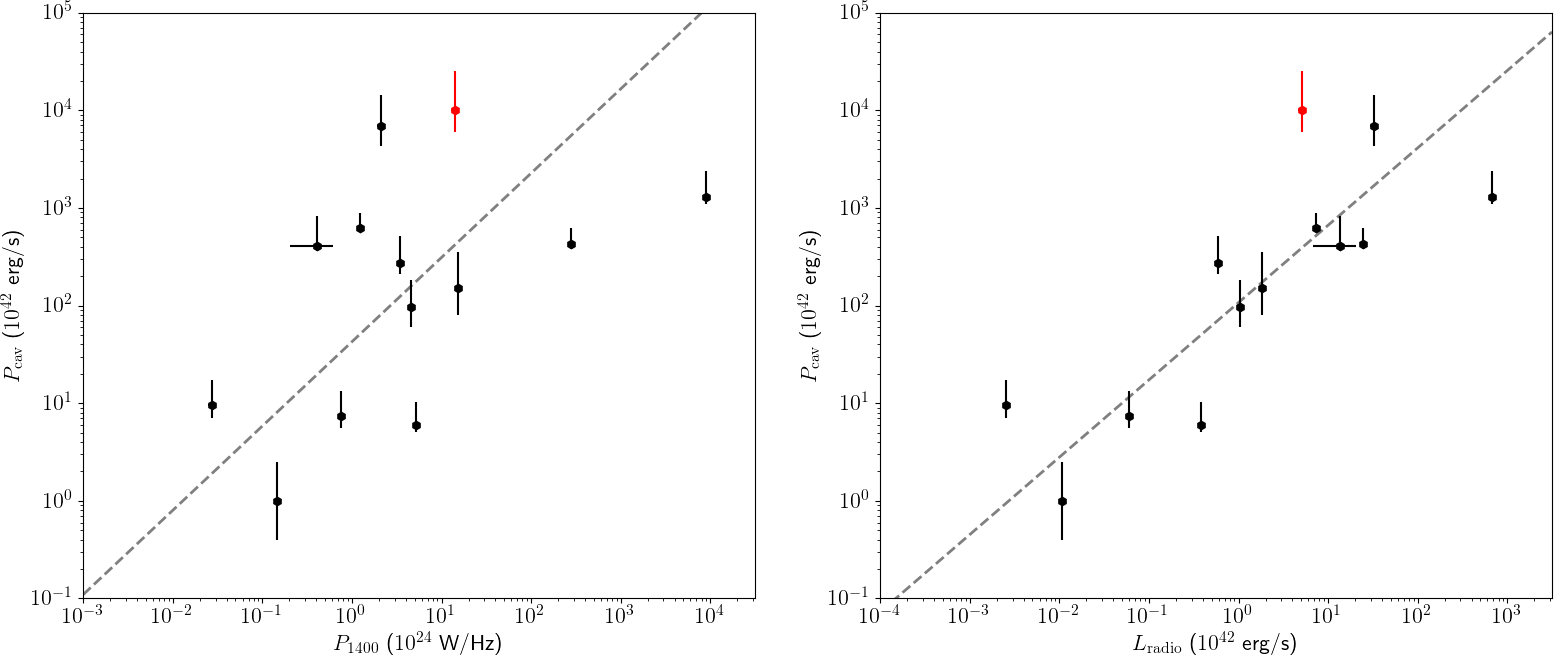}
    \caption{\textit{Left:} radio luminosity of the lobes at 1.4 GHz versus cavity power for the Phoenix cluster (red) and the sample of mini-halos from \citet{birzan08} (black). The dashed line shows the best power-law fit through the data, and is given by \(\log P_\text{cav} [10^{42} \text{erg/s}] = (0.86 \pm 0.20)\log P_\text{1400} [10^{24} \text{W/Hz}] + (1.63 \pm 0.29)\). \textit{Right:} bolometric radio luminosities of the lobes between 10 MHz and 10 GHz for the same sample. The dashed line shows the best power-law fit through the data, and is given by \(\log P_\text{cav} [10^{42} \text{erg/s}] = (0.79 \pm 0.15)\log L_\text{radio} [10^{42} \text{erg/s}] + (2.03 \pm 0.19)\). Data for the Phoenix cluster are from \citet{mcdonald19} and this work, and data for the other clusters are from \citet{birzan08}.}
    \label{fig:radiocav_corr}
\end{figure*}

\begin{figure}
    \centering
    \includegraphics[width=\columnwidth]{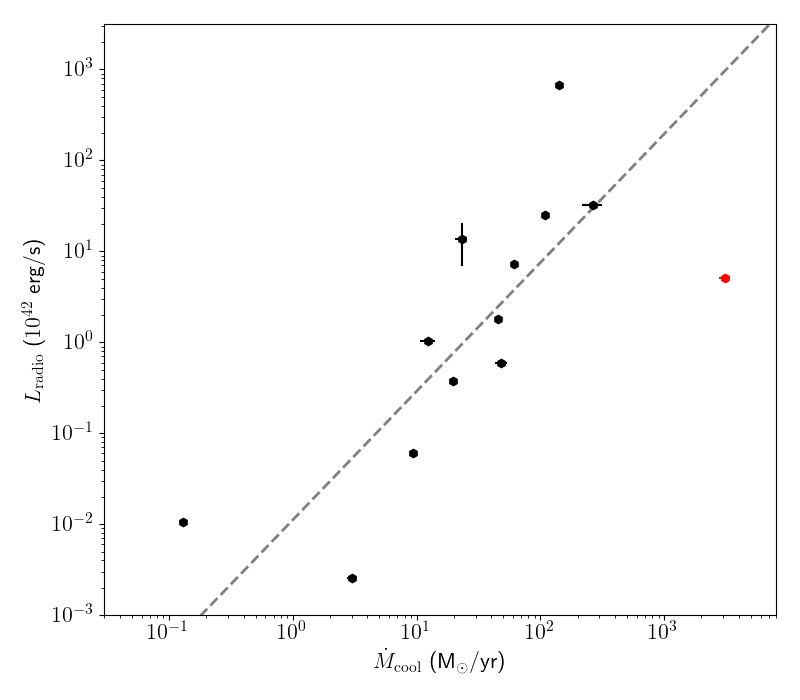}
    \caption{ICM cooling rate \(\dot{M}_\mathrm{cool}\) versus bolometric radio luminosity of the lobes between 10 MHz and 10 GHz. The red data point indicates the Phoenix cluster, and the black data points indicate the sample of \citet{birzan08}. The dashed line shows the best power-law fit through the data, and is given by \(\log L_\text{radio} [10^{42} \text{erg/s}] = (1.41 \pm 0.35)\log \dot{M}_\text{cool} [\text{M}_\odot / \text{yr}] - (1.95 \pm 0.53)\). Data for the Phoenix cluster are from \citet{mcdonald19} and this work, and data for the other clusters are from \citet{birzan08, mcdonald19}.}
    \label{fig:Mdot_Lbol}
\end{figure}

In addition, we can test predictions from hadronic models on the spectral index of the mini-halo. According to a hadronic model, the radio-emitting electrons are injected by collisions between cosmic-ray protons from the AGN and thermal protons in the ICM. This implies that the synchrotron spectral index \(\alpha\) is directly proportional to the cosmic ray proton injection spectral index \(\delta\) as \(\alpha \approx \delta/2\) \citep[e.g.][]{blasi99, pfrommer04, brunetti17}. Therefore, our spectral index of \(\alpha = -0.95 \pm 0.10\) requires a cosmic ray proton injection spectral index of \(\delta = -1.90 \pm 0.20\), but such a flat distribution is only marginally consistent with the generally observed values of \(\delta\) in the range of \(-2.1\) to \(-2.4\) \citep[e.g.][]{volk96, blasi99, schlickeiser02, ensslin03, pinzke10}.

As we find that both the magnetic field configurations obtained by assuming a hadronic model and the relatively flat spectral index of the mini-halo are inconsistent with the values generally reported in literature, we conclude that our results disfavor a pure hadronic origin of the radio emission, although we can not exclude that proton-proton collisions played a role in the origin of seed electrons for the re-acceleration. Therefore, a turbulent re-acceleration model is the preferred model to explain the origin of the mini-halo in the Phoenix cluster. However, the question remains as to what causes this turbulence.

\citet{mazzotta08} observed a correlation between the mini-halo emission and the spiral-shaped cold fronts produced by sloshing of the gas in the cool core \citep{markevitch03, ascasibar06}. Magneto-hydrodynamic simulations by \citet{zuhone11a, zuhone11b, zuhone13, zuhone15} show that this sloshing can induce turbulence and amplify the magnetic fields required to re-accelerate thermal electrons to relativistic speeds. In addition, their simulations predict luminosities and spectral indices that are in agreement with observations. Recent research by \citet{richardlaferriere20} suggests that in addition to sloshing, AGN feedback may also contribute significantly to the amount of turbulence. For the Phoenix cluster, we find that the extent of the mini-halo matches with the extent of the sloshing pattern in the ICM observed using X-ray data (see Figure \ref{fig:Xray_res_LXband_overlay}). Towards the south of the AGN, the mini-halo appears to be confined to the overdense region in the ICM, whereas towards the west, the mini-halo appears to be confined to the underdense region in the ICM.

On the other hand, \citet{gitti02,gitti04a, gitti07} suggest that the turbulence in the core can be induced by the strong cooling flow accreting onto the cool core. If so, a direct relation is expected between the mini-halo cooling flow power \(P_\mathrm{cf}\) and the integrated mini-halo radio power \(\nu P_\mathrm{MH}\). The cooling flow power is calculated as \(P_\mathrm{cf}=\dot{M}kT/\mu m_\mathrm{H}\), where \(\dot{M}\) is the mass accretion rate, \(k\) is Boltzmann's constant, \(T\) is the temperature of the ICM, \(\mu\) is the mean molecular weight and \(m_\mathrm{H}\) is the mass of a hydrogen atom. Based on the sample of mini-halos from \citet{giacintucci19}, we plot the correlation between these two properties, as shown in Figure \ref{fig:Pcf_PMH}. As previously verified by \citet{doria12} and \citet{bravi15} using different samples, this correlation between the cooling flow power and the integrated radio power of the mini-halo is indeed present. However, the relatively large intrinsic scatter suggests that this correlation may not indicate a causal connection between cooling flow rate and turbulence in the cool core. Instead, this correlation could, for example, emerge due to dependence on a third parameter.

Therefore, we conclude that the turbulent re-acceleration model powered by sloshing is the prime candidate to explain the origin of the mini-halo. Based on the close similarities between the mini-halo emission region and the sloshing region, and the agreement between the observed spectral index and theoretical predictions, we find sloshing to be the main source of turbulence in the ICM, although we do not exclude that the exceptionally strong cooling flow in the Phoenix cluster may have partly contributed to the turbulence.

\subsection{The extreme feedback in the Phoenix cluster}

Two of the main open questions about the feedback process in the Phoenix cluster remain why this cluster features such a strong cooling flow and why the radio and X-ray observations are so disconnected. According to the explanation recently proposed by \citet{mcdonald18}, the strong cooling flow could be a consequence of the way the most massive clusters (such as the Phoenix cluster) are formed. As lighter clusters merge and form more massive clusters, their low-entropy gas content will merge relatively fast compared to their central galaxies. This delayed merging of the central galaxies will lead to the most massive clusters having relatively underweight supermassive black holes powering their AGNs, as they generally have a rich and recent merging history. For supermassive black holes, it has been observed that the mechanical power levels off as the accretion rate reaches a few percent of the Eddington rate \citep{russell13}. This is in contrast to the radiative power of the AGN, which continues to increase with the accretion rate. In the context of a galaxy cluster, this means that if the central SMBH is underweight, the same accretion rate of the AGN is more likely to reach this saturation point, and result in the AGN being unable to compensate for the cooling flow with mechanical feedback. Verifying this explanation would require precise measurements of the black hole masses, which we can not derive from our radio observations. Therefore, definitively answering this open question is not the goal of this work. However, we can investigate if our observations are consistent with this explanation. 

According to this explanation, the overall feedback process in the Phoenix cluster is not abnormal. Instead, the Phoenix cluster should simply be in the tail of the distribution. 
To see in which aspects the Phoenix cluster is an outlier to the general population, we look into its cavity power and ICM cooling rate. The cavity power \(P_\mathrm{cav}\) of a cluster can be estimated by dividing the total enthalpy \(E_\mathrm{cav}\) of the cavities by the buoyant rise time. Here, the total enthalpy \(E_\mathrm{cav}\) can be calculated as \(E_\mathrm{cav}=4PV\), where \(P\) is the total pressure of the ICM at the position of the cavities and \(V\) is the volume of the cavities. The buoyant rise time is the time required for the cavities to move from the AGN to their present position under the assumption that they rise through the ICM buoyantly. The ICM cooling rate within a particular radius indicates how much mass in the ICM cools down per year due to the emission of thermal bremsstrahlung, and can be calculated as the ICM mass enclosed within the given radius divided by the cooling time. We plot the cavity power versus the ICM cooling rate for the Phoenix cluster and the sample of clusters from \citet{rafferty06}, as shown in Figure \ref{fig:Mdot_Pcav}. Here we find that the Phoenix cluster is the most extreme cluster in the sample, both in terms of the cooling flow rate and the cavity power. However, the Phoenix cluster is consistent with the observed correlation. Similarly, we have plotted the radio luminosity of the jetted outflows at 1.4 GHz, as well as the bolometric radio luminosity, against the cavity power, for a sample of clusters, as shown in Figure \ref{fig:radiocav_corr}. This is where the Phoenix cluster does appear to be an outlier, as the cavity power is too high for both the bolometric radio luminosity of the lobes and their radio luminosity at 1.4 GHz. This is amplified if we plot the cooling flow rate against the bolometric radio luminosity of the lobes, as shown in Figure \ref{fig:Mdot_Lbol}. In this comparison, the Phoenix cluster is a strong outlier in the correlation. For its cooling flow rate, the radio lobes in the Phoenix cluster are far too faint.

We observe a disconnection between the radio- and the X-ray properties of the Phoenix cluster. In the radio regime, the AGN in the Phoenix cluster appears to be relatively modest, whereas in the X-ray regime, both the AGN and the cooling flow are among the strongest of all known clusters. We suspect that this difference between the radio and the X-ray may be caused by strong time variability of the AGN on Myr-timescales. In our radio observations, both the northern and southern lobes clearly appear to be detached from the AGN, and this feature is also visible in the ATCA observations presented by \citet{akahori20}. In addition, we find that the central AGN is not a perfect point source, but it is elongated towards the northwest and southeast, indicating that new outflows are being detected. 
Outburst power has been observed to vary from factors of several in Hydra A \citep{wise07} to two orders of magnitude in MS0735 \citep{vantyghem14} over timescales of tens of Myr. This observed variability has been corroborated by numerical simulations \citep[e.g.][]{li15, prasad15}, providing an explanation for the lobe structure in the Phoenix cluster. It is possible that the sloshing in the ICM contributes in part to this AGN variability by displacing the accretion material of the AGN, similar to as observed in A2495 by \citet{pasini19}. However, whereas the BCG in A2495 appears to be oscillating back and forth through the cool core, the BCG in the Phoenix cluster is stationary at the center of the cool core. This means that sloshing can only affect the AGN activity if it inhibits the cooling flow. Detailed simulations by \citet{zuhone10} suggest that sloshing mainly introduces variability in the cooling flow on timescales of a Gyr or more. These timescales are too long to explain the AGN variability in the Phoenix cluster, so we consider it to be unlikely that sloshing plays a major role. In this scenario, the lack of radio luminosity would be caused by a break in the AGN activity, as the only contribution to the lobe radio luminosity is caused by a relatively short and old outburst. This break in the AGN activity would not necessarily affect the currently observed cavity power, as the cavities require time to expand, and are therefore more dependent on a later stage of an outburst. However, the current volume of the cavities will not be sustainable due to this variability of the AGN. As the break in the outflows reaches the cavities, they will deflate, resulting in the average cavity power likely being lower than the presently observed value. 

Finally, the relatively low amount of jet precession observed in the Phoenix cluster by \citet{akahori20} and our X-band observations likely also contributes to the low ICM re-heating efficiency of the mechanical feedback, as the energy from the AGN is not distributed isotropically, but rather predominantly along the direction of the jets.

The explanation by \citet{mcdonald18} combined with time variability of the AGN activity and a low jet precession angle may resolve some of the remaining open questions on the AGN feedback in the Phoenix cluster, although a more quantitative investigation would be required to confirm the validity of this model.

\section{Conclusions}

In this paper, we have presented new \textit{Karl G. Jansky Very Large Array} observations, enabling the radio lobes of the AGN and the mini-halo in the Phoenix cluster to be studied in detail at frequencies from 1 to 12 GHz. In particular, our observations resolve the radio lobes of the AGN in all four frequency bands, and the mini-halo can be detected in both our L- and S-bands. Using these multifrequency observations, we have studied the remarkable feedback scenario in the Phoenix cluster and the origin of its mini-halo.

We find that the total flux density of the source at 1.5 GHz is 33.8 \(\pm\) 1.7 mJy, with an overall spectral index of \(-1.12 \pm 0.02\). Using our L- and S-band observations, we find that the mini-halo has an average spectral index of \(-0.95 \pm 0.10\). By subtracting compact emission and through radial profile fitting, we find that the mini-halo has a total flux density at 1.5 GHz of \(8.5 \pm 0.9\) mJy, which corresponds to a radio luminosity at 1.4 GHz of \((13.0 \pm 1.4)\times10^{24}\ \mathrm{W}\ \mathrm{Hz}^{-1}\). In addition, we find that the mini-halo has a maximum observable radius at L-band of 240 kpc. At 1.5 GHz, the radio lobes show a total flux density of \(7.6 \pm 0.8\) mJy, and spectral indices with respect to 10 GHz of \(-1.35 \pm 0.07\) (northern lobe) and \(-1.30 \pm 0.12\) (southern lobe).

Due to the relatively flat spectral index of the mini-halo, the low index \(k\) and the extreme magnetic field configuration we obtain by assuming a hadronic model, we conclude that our results disfavor a pure hadronic origin of the mini-halo. On the contrary, we confirm the correlation between the cooling flow power and the radio luminosity of the mini-halo with respect to other clusters. Also, we observe the sloshing pattern in the ICM to match with the extent of the mini-halo at radio frequencies, and find our integrated spectral index of the mini-halo to be consistent with numerical predictions for a turbulent re-acceleration model. For these reasons, we conclude that a turbulent re-acceleration model is the preferred model to explain the origin of the mini-halo, and in particular favor sloshing in the cool-core as an explanation for the turbulence.

By measuring the flux density of the radio lobes in L-band for the first time, the mechanical feedback in the Phoenix cluster has been studied from a radio perspective. We observe a disconnection between the X-ray and radio properties of the Phoenix cluster: the cavity power and cooling flow rate are both among the most extreme ever measured, whereas the bolometric radio luminosity of the lobes is relatively modest. We find that the feedback in the Phoenix cluster is overall consistent with the proposed explanation by \citet{mcdonald18}, which states that the strong cooling rate and inefficient feedback are characteristic for more massive clusters. Strong time variability of the AGN activity on Myr-timescales may explain the disconnection between the radio and the X-ray properties of the system. Finally, a small amount of jet precession likely also contributes to the low ICM re-heating efficiency of the AGN feedback.

For future research, it would be valuable to obtain more high-resolution radio observations of a sample of mini-halos to further test the correlation between the mini-halo emission and the sloshing pattern in the ICM, as this could provide very strong evidence for the turbulent re-acceleration model, and current results look promising. In addition, low-frequency observations may test if sloshing of the ICM produces more extended, ultra-steep spectrum emission beyond the cold fronts, as observed in the case of PSZ1G139.61+24 and RXJ1720.1+2638 by \citet{savini18, savini19}.

\begin{acknowledgements}
      We would like to thank the anonymous referee for useful comments. RT and RJvW acknowledge support from the ERC Starting Grant ClusterWeb 804208. Support for this work was provided to MM by NASA through Chandra Award Number GO7-18124 issued by the Chandra X-ray Observatory Center, which is operated by the Smithsonian Astrophysical Observatory for and on behalf of the National Aeronautics Space Administration under contract NAS8-03060.
\end{acknowledgements}

\bibliographystyle{aa}
\bibliography{refs}

\end{document}